\documentclass[aps,groupaddress,floats,preprint,showpacs]{revtex4}
\usepackage{amsfonts}
\usepackage{amssymb}
\usepackage{bm}
\usepackage[final]{graphicx}

\begin{document}
\bibliographystyle{revtex}


\newcommand{\alps}{\ensuremath{\alpha_s}}
\newcommand{\qbar}{\bar{q}}
\newcommand{\beq}{\begin{equation}}
\newcommand{\eeq}{\end{equation}}
\newcommand{\beqa}{\begin{eqnarray}}
\newcommand{\eeqa}{\end{eqnarray}}
\newcommand{\mq}{m_Q}
\newcommand{\mn}{m_N}
\newcommand{\bb}{\langle}
\newcommand{\kb}{\rangle}
\newcommand{\st}{\ensuremath{\sqrt{\sigma}}}
\newcommand{\rvec}{\mathbf{r}}
\newcommand{\bvec}[1]{\ensuremath{\mathbf{#1}}}
\newcommand{\bra}[1]{\ensuremath{\bb#1|}}
\newcommand{\ket}[1]{\ensuremath{|#1\kb}}
\newcommand{\gft}{\ensuremath{\gamma_{FT}}}
\newcommand{\bfsig}{\mbox{\boldmath{$\sigma$}}}
\newcommand{\bfnab}{\mbox{\boldmath{$\nabla$}}}
\newcommand{\bftau}{\mbox{\boldmath{$\tau$}}}
\newcommand{\spup}{\uparrow}
\newcommand{\spd}{\downarrow}
\newcommand{\hbarom}{\frac{\hbar^2}{m_Q}}
\newcommand{\vnn}{\ensuremath{\hat{v}_{NN}}}
\newcommand{\argonne}{\ensuremath{v_{18}}}
\newcommand{\lqcd}{\ensuremath{\mathcal{L}_{QCD}}}
\newcommand{\lgf}{\ensuremath{\mathcal{L}_g}}
\newcommand{\lqm}{\ensuremath{\mathcal{L}_q}}
\newcommand{\lqg}{\ensuremath{\mathcal{L}_{qg}}}
\newcommand{\nn}{\ensuremath{NN}}
\newcommand{\hpnd}{\ensuremath{H_{\pi N\Delta}}}
\newcommand{\hpqq}{\ensuremath{H_{\pi qq}}}
\newcommand{\fpnn}{\ensuremath{f_{\pi NN}}}
\newcommand{\fpnd}{\ensuremath{f_{\pi N\Delta}}}
\newcommand{\fpqq}{\ensuremath{f_{\pi qq}}}
\newcommand{\ylm}{\ensuremath{Y_\ell^m}}
\newcommand{\ylmc}{\ensuremath{Y_\ell^{m*}}}
\newcommand{\qbh}{\hat{\bvec{q}}}
\newcommand{\xbh}{\hat{\bvec{X}}}
\newcommand{\dt}{\Delta\tau}
\newcommand{\qmag}{|\bvec{q}|}
\newcommand{\pmag}{|\bvec{p}|}
\newcommand{\oas}{\ensuremath{\mathcal{O}(\alpha_s)}}
\newcommand{\vtxb}{\ensuremath{\Lambda_\mu(p',p)}}
\newcommand{\vtxp}{\ensuremath{\Lambda^\mu(p',p)}}
\newcommand{\pwqp}{e^{i\bvec{q}\cdot\bvec{r}}}
\newcommand{\pwqm}{e^{-i\bvec{q}\cdot\bvec{r}}}
\newcommand{\gsa}[1]{\ensuremath{\bb#1\kb_0}}
\newcommand{\oer}[1]{\mathcal{O}\left(\frac{1}{\qmag^{#1}}\right)}
\newcommand{\nub}[1]{\overline{\nu^{#1}}}
\newcommand{\yjm}{\mathcal{Y}_{jm}}
\newcommand{\balph}{\mbox{\boldmath{$\alpha$}}}
\newcommand{\bgam}{\mbox{\boldmath{$\gamma$}}}

\title
{Final state interaction contribution to the response of confined
relativistic particles}

\author{Mark W.\ Paris}
\email[]{paris@lanl.gov}
\affiliation{Theoretical Division,
Los Alamos National Laboratory, Los Alamos, New Mexico 87545,}
\affiliation{Department of Physics,
University of Illinois at Urbana-Champaign,
1110 West Green Street, Urbana, Illinois 61801}
\author{Vijay R.\ Pandharipande}
\email[]{vrp@uiuc.edu}
\affiliation{Department of Physics,
University of Illinois at Urbana-Champaign,
1110 West Green Street, Urbana, Illinois 61801}

\date{\today}

\begin{abstract}
\medskip

We report studies of the response of a massless particle confined
by a potential. At large momentum transfer $\bvec{q}$ it exhibits
$\tilde{y}$ or equivalently Nachtmann $\xi$ scaling, and acquires a
constant width independent of $\bvec{q}$.  This width has a contribution
from the final state interactions of the struck particle, which
persists in the $\qmag\rightarrow\infty$ limit.  The width of the
response predicted using plane wave impulse approximation is smaller
because of the neglect of final state interactions in that approximation.
However, the exact response may be obtained by folding the approximate
response with a function representing final state interaction effects.
We also study the response obtained from the momentum distribution
assuming that the particle is on the energy shell both before
and after being struck. Quantitative results are presented for
the special case of a linear confining potential. In this case the
response predicted with the on-shell approximation has correct
values for the total strength, mean energy and width, however
its shape is wrong.

\end{abstract}
\maketitle

\section{Introduction}

Scattering of high energy probes from composite systems, such as electron
scattering by nuclei \cite{Frois91} and nucleons \cite{ESW},
or neutron scattering by liquid helium \cite{SS89}, is often used to
study the structure of the bound system. The common assumption is that
in deep inelastic scattering (DIS) at sufficiently high energy 
the probe is incoherently scattered by the constituents of the
system. In the plane wave impulse approximation (PWIA), which neglects
final state interaction (FSI) effects, DIS is directly related to
the momentum and energy distribution of the constituents in the
target.

The role of FSI effects has been studied extensively in
electron scattering from nuclear targets \cite{BFFMPS91,BP93}
and neutron scattering from liquid helium \cite{SS89}. Recently
it has been suggested that they may also influence DIS of
leptons by hadrons \cite{Brodsky}.
In the present study we focus on scattering from targets with
confined constituents. The corresponding physical case concerns
DIS from nucleons where, in distinction from the nuclear and
liquid helium cases, the constituents are confined in both the
initial and final states.

We \cite{Paris01} have recently studied the response of a
massless particle confined by a linear (flux-tube) potential
using the simple model Hamiltonian
\beq
\label{eqn:H}
H = \sqrt{|{\bvec{p}}|^2} + \st r,
\eeq
to gain insights into DIS of leptons by
hadrons.  The response to a hypothetical scalar probe is calculated
as a function of the momentum and energy transfer, $\bvec{q}$,
$\nu$ in the lab frame. It is given by
\beq
\label{eqn:Rx}
R(\bvec{q},\nu) = \sum_I |\bra{I} e^{i \bvec{q}\cdot\rvec}
\ket{0}|^2 \delta(E_I - E_0 - \nu),
\eeq
where the sum is over all energy eigenstates $I$.
The natural scaling variable in the many-body theory approach to
DIS is $\tilde{y}=\nu-|\bvec{q}|$ \cite{BPS00}. In the large 
$|\bvec{q}|$ or scaling limit the response depends only on
$\tilde{y}$, not on $\qmag$ and $\nu$ independently.
This scaling of the response is equivalent to the
Nachtmann scaling, since the Nachtmann \cite{ON} scaling variable
$\xi = -\tilde{y}/M$,
where $M$ is the hadron mass. 
In Ref.\ \cite{Paris01} the $R(\bvec{q},\nu)$ is calculated for
$|\bvec{q}| \leq 10 $ GeV for the typical value of the string
tension $\sqrt{\sigma} = 1$ GeV/fm in QCD,
by calculating all the relevant states $|I\rangle$
contributing to the response.

In this work we study the effects of the
FSI of the struck particle on the response;
analytic calculations of the width of the response are
presented in Sec.\ref{sec:anl},  and the numerical results
for a linear confining potential are  given
in Sec.\ref{sec:num}.  Both indicate that the FSI increase the
width of the response beyond that predicted by PWIA. 
The analytic calculations also consider
the nonrelativistic problem, in which $\bvec{q}$ is large compared to
all the momenta in the target, but smaller than the constituent
mass $m$.
The main differences between the
nonrelativistic and the relativistic response are that the
former peaks at $\nu=\qmag^2/2m$ and has a width proportional
to $|\bvec{q}|$, while the latter peaks at $\nu \sim |\bvec{q}|$,
and has a constant width in the scaling limit.  The folding
function \cite{OB} representing the effects of the FSI on the
response is discussed in the last section, Sec.\ref{sec:cncl},
where conclusions are given.

\section{Moments of the Response}
\label{sec:anl}
In the case of a single confined particle, the state of the
system after the probe has struck the target is
\beq
\label{eqn:X}
\ket{X} = \pwqp\ket{0} ,
\eeq
where $\ket{0}$ denotes the ground state of the particle. The state
$\ket{X}$ is not an eigenstate of the Hamiltonian and therefore has
a distribution in energy. It  has a unit norm,
$\bb X|X \kb = \bra{0}\pwqm\pwqp\ket{0} = 1$.
The total strength of the response, given by the static
structure function
\beq
S(\qmag) = \int_0^{\infty} d\nu\, R(\qmag,\nu),
\label{eqn:S}
\eeq
is therefore unity.
In many-body systems $S(\qmag)$ is not necessarily equal to one.
Subsequent formulas pertain to the general case
and show factors of $S(\qmag)$ explicitly.

The mean
excitation energy of the state $\ket{X}$ is given by the first
moment of the response:
\beq
\label{eqn:barnu}
\overline{\nu}(\qmag) = \frac{1}{S(\qmag )} \bra{X} H-E_0 \ket{X}
=\frac{1}{S(\qmag)}\int_0^\infty d\nu\,\nu\, R(\qmag,\nu).
\eeq
The width of the distribution in energy is characterized
by the second moment of the energy about the mean:
\beq
\label{eqn:delta}
\Delta^2(\qmag) = \frac{1}{S(\qmag)}\bra{X}
\left( H - \frac{\bra{X}H\ket{X}}{S(\qmag)} \right)^{\!2} \ket{X}.
\eeq

For example, if the response is Gaussian, it
is completely determined by these three moments:
\beq
\label{eqn:Rg}
R_G(\bvec{q},\nu) = \frac{S(\qmag)}{\sqrt{\pi\Delta^2(\qmag)}}
e^{-(\nu - \overline{\nu}(\qmag))^2/\Delta^2(\qmag)}.
\eeq
The full-width at
half-maximum (FWHM) of a Gaussian response is given by
$2\Delta(\qmag)\sqrt{\ln 2} \approx\frac{5}{3}\Delta(\qmag)$.
Obviously many more moments are required to describe a response
of more general shape.  Nevertheless, a
necessary condition for the occurrence of the $\tilde{y}$
or Nachtmann $\xi$ scaling is that $S(\qmag)$,
$\overline{\nu}(\qmag)-\qmag$, $\Delta(\qmag)$, and all the
higher moments become independent of $\qmag$, since the
response depends only on $\nu - \qmag$ when $\qmag \rightarrow \infty$.

In the remainder of this section we will calculate
the average excitation energy, $\overline{\nu}(\qmag)$
and the width, $\Delta(\qmag)$ for the model Hamiltonian of
Eq.(\ref{eqn:H}) in the scaling limit, $\qmag\rightarrow\infty$.
They are calculated analytically for a general
potential $V(r)$. Section \ref{sec:num} presents numerical
results for the case of a linear confining potential. 
This analysis allows the comparison of the results
for the exact response with the following two common
approximations. The PWIA,
which assumes that the final state of the struck particle
can be approximated by a plane wave in an average potential, is
commonly used in nuclear physics. 
And the on-shell approximation (OSA), in which the
struck particle is assumed to be in a plane wave state
with the energy of that of a free particle, before and after
the interaction with the probe, used in hadron physics. 

We proceed with the evaluation of the average excitation energy
[Eq.(\ref{eqn:barnu})] for the case of the exact response.
The kinetic energy term of the
matrix element $\bra{X}H\ket{X}$ is
\beq
\label{eqn:kx}
\bra{X}\sqrt{|\bvec{p}|^2}\ket{X} = \int \frac{d^3k}{(2\pi)^3} n(k)
|\bvec{k}+\bvec{q}|,
\eeq
where we have transformed to momentum space and $n(k)
= |\psi_0(k)|^2$. In the scaling limit, with
$\bvec{q}=\qmag\hat{\bvec{z}}$, we can expand $|\bvec{q}+\bvec{k}|$
and obtain:
\beq
\label{eqn:kpq}
|\bvec{k}+\bvec{q}| = \qmag + k_z + \frac{1}{2\qmag}k^2_\perp
- \frac{1}{2\qmag^2}k^2_\perp k_z
+ \oer{3}.
\eeq
Here $\oer{3}$ denotes the neglected terms of that and higher order,
and $k^2_\perp = k^2_x + k^2_y$. Substituting this
expansion into Eq.(\ref{eqn:barnu}) yields,
\beq
\label{eqn:barnux}
\overline{\nu}(\qmag) = \qmag + \gsa{V} - E_0 + \frac{1}{3\qmag}\gsa{k^2}
+ \oer{3};
\eeq
the term of $\oer{2}$ is zero by symmetry.
Note that $E_0-\gsa{V}=\gsa{T}$, where $T$ denotes the kinetic energy.
Thus $\overline{\nu}(\qmag)= \qmag - \gsa{T}$ in the limit $\qmag 
\rightarrow \infty$.
The requirement that $\overline{\nu}(\qmag)-\qmag$ becomes constant
is naturally satisfied in this limit.

The width [Eq.(\ref{eqn:delta})] depends on  the matrix element
$\bra{X}H^2\ket{X}$;
$H^2 = p^2 + \sqrt{p^2}V(r) + V(r)\sqrt{p^2} + V^2(r)$.
The first term of $H^2$ gives:
\beq
\bra{X} p^2 \ket{X}=\gsa{|\bvec{k}+\bvec{q}|^2} = \qmag^2 + \gsa{k^2} \ .
\eeq
The average of the second and third terms in $H^2$ may be shown
to be equal to each other and
\beqa
\label{eqn:ct}
\bra{X}\sqrt{p^2}V\ket{X}
&=& \int\frac{d^3k}{(2\pi)^3} \psi_0^*(k) |\bvec{q}+\bvec{k}| [V\psi_0](k)
\nonumber \\
&\approx& \qmag\gsa{V} + \frac{1}{2\qmag} \gsa{k^2_\perp V}
+ \oer{3},
\eeqa
where $[V\psi_0](k)
= \int d^3r e^{i\bvec{k}\cdot\bvec{r}} V(r)\psi_0(r)$. Substitution into
Eq.(\ref{eqn:delta}) with $S(\qmag)=1$ gives:
\beq
\label{eqn:delgs}
\Delta^2(\qmag) = \frac{1}{3}\gsa{k^2} + \gsa{V^2} - \gsa{V}^2
+ \frac{2}{3\qmag}\left(\gsa{k^2V}-\gsa{k^2}\gsa{V}\right)
+ \oer{2}.
\eeq
This expression demonstrates that the width of the exact response
is independent of $\qmag$ in the limit $\qmag\rightarrow\infty$
as necessary for $\tilde{y}$ scaling.
It also shows that the width
has a kinematic contribution dependent upon the target
momentum distribution, and an
additional interaction contribution.

As mentioned, the PWIA assumes that a constituent of momentum $\bvec{k}$,
after being struck by the probe, may be described
by a plane wave with momentum $\bvec{k+q}$ in an assumed
average potential chosen to give the exact $\overline{\nu}$
of Eq.(\ref{eqn:barnux}). The PWIA response is
\beq
\label{eqn:Rpw}
R_{PWIA}(\qmag,\nu) = \int \frac{d^3k}{(2\pi)^3} n(k)
\delta( e(\bvec{k}+\bvec{q}) - E_0 - \nu )
\eeq
where $e(\bvec{k}+\bvec{q})$ is the energy of the plane wave,
taken to be:
\beq
\label{eqn:epw}
e(\bvec{k}+\bvec{q}) = |\bvec{k}+\bvec{q}| + \gsa{V}.
\eeq
The first term is the kinetic energy of the struck particle.
The second term is the average of the potential in the
final state of the system given by, $\bra{X}V\ket{X}=
\bra{0}\pwqm V\pwqp\ket{0}=\gsa{V}$.
The $R_{PWIA}(\qmag,\nu)$ exhibits $\tilde{y}$ scaling as can
be easily seen by expanding the argument of the $\delta$ function
in $R_{PWIA}$.  In the large $\qmag$ limit $R_{PWIA}$ depends
only on $\tilde{y}=\nu-\qmag$.

The average energy and the width of the PWIA response are
calculated using:
\beq
\label{eqn:mr}
\int_0^\infty d\nu\, \nu^n R_{PWIA}(\qmag,\nu)
= \int_0^\infty d\nu\, \nu^n
\int \frac{d^3k}{(2\pi)^3} n(k)
\delta( e(\bvec{k}+\bvec{q}) - E_0 - \nu ).
\eeq
The average excitation energy, obtained
with $n=1$, agrees with the exact result in Eq.(\ref{eqn:barnux})
by construction.

The width of the PWIA response, however, is not the same as the
width of the exact response.
\beq
\label{eqn:delpw}
\Delta^2_{PWIA} = \frac{1}{3}\gsa{k^2}
+ \oer{2}
\eeq
contains only the first term of the exact result [Eq.(\ref{eqn:delgs})]
due to the target momentum distribution.  The second term,  
$\gsa{V^2}-\gsa{V}^2$, of the exact $\Delta^2$ represents
the FSI contribution neglected in the PWIA.  It does not vanish in the
$|\bvec{q}| \rightarrow \infty$ limit for relativistic kinematics.

In the non-relativistic case,
$H_{NR} = \frac{|\bvec{p}|^2}{2m} + V(r)$,
the exact $\overline{\nu}$
is given by:
\beq
\label{eqn:barnupwnr}
\overline{\nu}_{NR} = \frac{1}{2m}\qmag^2 \ .
\eeq
The nonrelativistic-PWIA (NR-PWIA) response:
\beq
\label{eqn:Rpwnr}
R_{NR-PWIA}(\qmag,\nu) = \int \frac{d^3k}{(2\pi)^3} n(k)
\delta\left( \frac{|\bvec{q}+\bvec{k}|^2}{2m} + \gsa{V} - E_0 - \nu \right)
\eeq
gives the exact $\overline{\nu}$.
For the width of the NR-PWIA response we obtain:
\beq
\label{eqn:delpwnr}
\Delta^2_{NR-PWIA}(\qmag) = \qmag^2 \frac{\gsa{k^2}}{3 m^2}
 + \frac{1}{4m^2} \left( \gsa{k^4} - \gsa{k^2}^2 \right).
\eeq
Note that in Eqs.(\ref{eqn:barnupwnr}) and (\ref{eqn:delpwnr})
we have {\em not} taken the $\qmag\rightarrow\infty$ limit.

The width of the exact NR response is:
\beq
\label{eqn:delnr}
\Delta^2_{NR}(\qmag) = \Delta^2_{NR-PWIA}
+ \frac{1}{m}\left(\gsa{Vk^2} - \gsa{V}\gsa{k^2}\right)
+ \gsa{V^2} - \gsa{V}^2 \ .
\eeq
It differs from $\Delta_{NR-PWIA}$ in terms of
order $1/|\bvec{q}|$ which can be neglected in the
scaling limit.  Thus, in contrast to the relativistic case,
the FSI do not increase
the width of the NR-PWIA response at large $|\bvec{q}|$.

Finally we consider the on-shell approximation (OSA) in
which the energy of
the struck constituent is that of a free relativistic
particle before and after the interaction with probe,
as assumed in the quark-parton model. The response in
OSA is
\beq
\label{eqn:Rosa}
R_{OSA}(\qmag,\nu)=\int\frac{d^3k}{(2\pi)^3} n(k)
\delta\left( |\bvec{k}+\bvec{q}|-|\bvec{k}|-\nu \right);
\eeq
it depends only on the momentum distribution of target constituents
and obeys $\tilde{y}$ scaling.
The average excitation in OSA is
\beq
\label{eqn:barnuosa}
\overline{\nu}_{OSA} = \qmag -\gsa{T} + \frac{1}{3\qmag}\gsa{k^2}
+\oer{3},
\eeq
and the width is given by:
\beq
\label{eqn:delosa}
\Delta^2_{OSA} = \frac{1}{3}\gsa{k^2} + \gsa{k^2}-\gsa{T}^2
+ \oer{}.
\eeq
The exact value of $\overline{\nu}$ [Eq.(\ref{eqn:barnux})]
is reproduced by the OSA for any potential.  However, the
$\Delta^2_{OSA}$ has $\gsa{k^2}-\gsa{T}^2$ in place of the
$\gsa{V^2}-\gsa{V}^2$ in the leading term of the exact $\Delta^2$
[Eq.(\ref{eqn:delgs})].
For a massless particle in a linear confining potential,
{\em i.e.}\ for the Hamiltonian of Eq.(\ref{eqn:H}),
$\gsa{T}=\gsa{V}$, and $\gsa{k^2}=\gsa{V^2}$. Therefore for
this particular Hamiltonian the OSA reproduces the
exact value of $\Delta$.

The calculated responses shown in the next section however indicate
that the shape of the OSA response is incorrect. We have calculated
the third moment of the $H$ in the state $\ket{X}$ and observed that
the exact and OSA responses indeed give different values. We expect the
moments to differ also for orders higher than the third since
the shapes of the exact and the OSA response are very different.

\begin{figure}
\includegraphics[ width=275pt, keepaspectratio, angle=0, clip ]{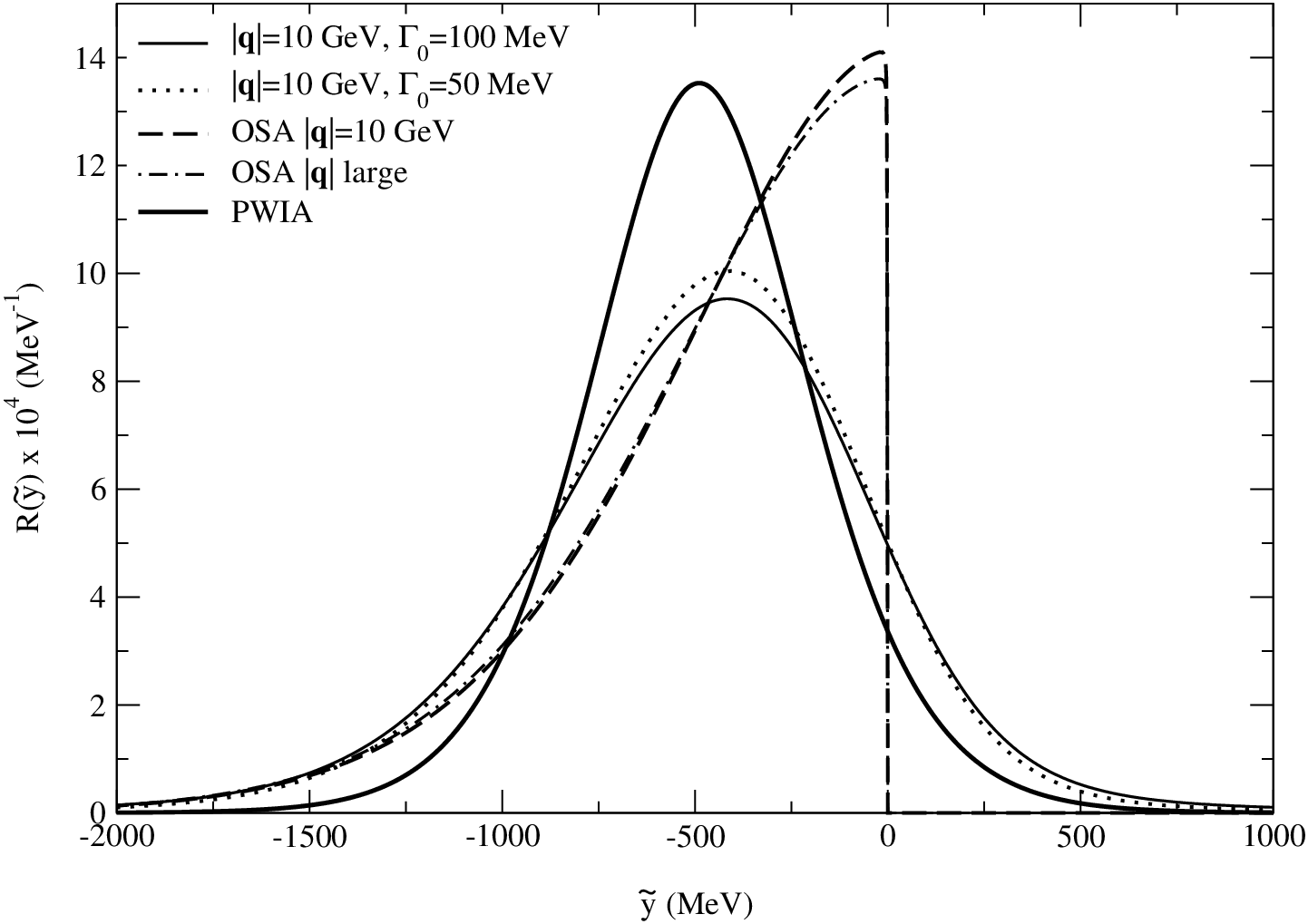}
\caption{\label{fig:rcf} The response versus $\tilde{y}$ calculated
exactly for $\Gamma_0=100$ MeV (thin solid curve) and $\Gamma_0=50$ MeV
(dotted curve). The response in OSA are shown for $\qmag=10$ GeV (dashed)
and $\qmag\rightarrow\infty$ (dot-dashed). The PWIA response for
$\qmag=10$ GeV and $\qmag\rightarrow\infty$ lie on essentially the
same (thick solid) curve.}
\end{figure}

\section{Numerical results}
\label{sec:num}

We first compare the response functions for $\qmag = 10 $ GeV
before comparing their moments. In Ref. \cite{Paris01} it has been
shown that the scaling limit is obtained for such values of
$\qmag$.  The exact response,
Eq.(\ref{eqn:Rx}), is a sequence of $\delta$ functions at
$\nu=E_I-E_0$. In order to obtain a smooth response
we assume decay widths $\Gamma_0$ for all the excited states. 
Note that the energies of the states $|I\rangle$ that contribute to
the response at $\qmag = 10 $ GeV are large, therefore their decay widths are not
affected by the energy dependent terms assumed in Ref. \cite{Paris01}.
The response including decay widths is given by:
\beq
\label{eqn:smear}
R(\bvec{q},\nu) = \sum_I|\langle I| e^{i \bvec{q} \cdot \bvec{r}}
|0 \rangle|^2 \left(\frac{\Gamma_0}{2\pi}\right) \frac{1}
{(E_I-E_0-\nu)^2 + \Gamma_0^2/4} \ .
\eeq
The responses obtained with $\Gamma_0=100$ and 50 MeV are shown in
Fig.\ \ref{fig:rcf}, along with the PWIA
and OSA responses for $\qmag = 10$ GeV and for $\qmag \rightarrow \infty$. 
The difference between the exact responses for $\Gamma_0=100$ and 50
MeV are much smaller than those between the exact and the approximate.

We note that the shape of the
PWIA response is qualitatively similar to that of the exact,
however, its width is too small.
This is a direct consequence of the neglect of interaction
terms in $\Delta$ [Eq.(\ref{eqn:delgs})] 
as discussed in the last section.  The width $\Delta$ of the response
for $\Gamma_0 \rightarrow 0$, is 409 MeV, while the $\Delta_{PWIA} = 326$
MeV.  Note that for a Gaussian response, Eq.(\ref{eqn:Rg})
the FWHM is $\sim 5 \Delta /3$.  The FWHM of the exact and the PWIA responses
shown in Fig.\ \ref{fig:rcf} are larger because they are not exactly Gaussian.

The OSA results in the discontinuous curves shown in
Fig.\ \ref{fig:rcf}. They are  discontinuous at the
lightline ($\qmag=\nu$) because the response of free
particles is limited to the spacelike region $\nu<\qmag$.
The discontinuity at $\tilde{y}=0$ is in clear conflict with
the exact response which is continuous across the lightline
and is non-zero in the timelike ($\tilde{y}>0$) region.
Therefore the OSA appears to be unsatisfactory even though for
the special case of a linear potential it
has the exact values of $S(\qmag)$, $\overline{\nu}(\qmag)$ and
$\Delta(\qmag)$.

\begin{figure}
\includegraphics[ width=275pt, keepaspectratio, angle=0, clip ]{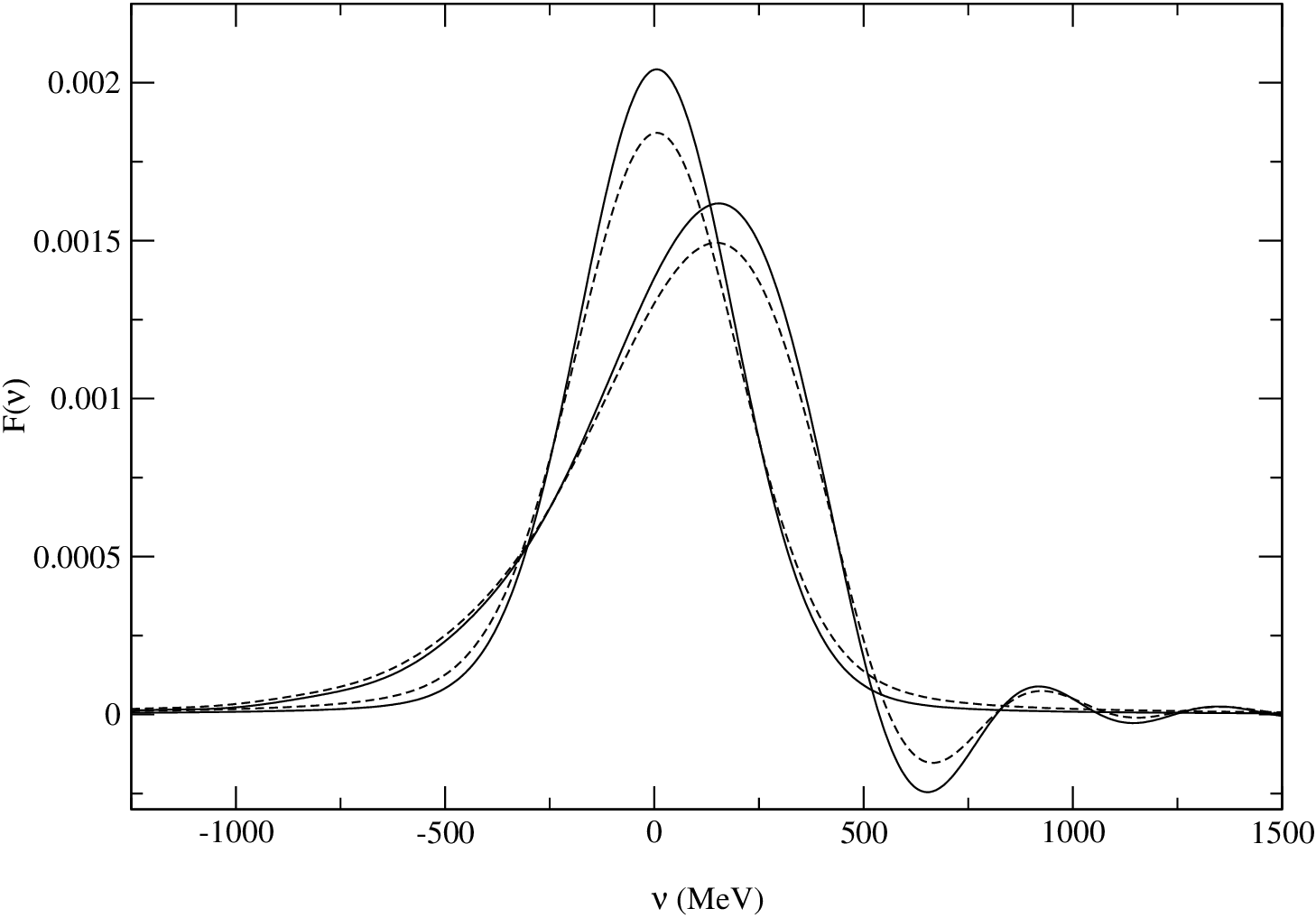}
\caption{\label{fig:ff}Folding functions for large $\qmag$. The solid
and dashed curves correspond to $\Gamma_0=50$ and 100 MeV, respectively.
The approximate folding functions $F_A$ of Eq.(\ref{eqn:aff}) are peaked
at $\nu=0$ while the folding functions obtained from the exact and PWIA
response by deconvolution are peaked at $\nu\sim 0.2$ GeV.}
\end{figure}

\section{FSI Folding function}
\label{sec:cncl}

The exact response may be obtained from $R_{PWIA}(\qmag,\nu)$
via convolution with the folding function \cite{SS89,BFFMPS91,OB}.
The states $\ket{\bvec{k}+\bvec{q}}$ used in the
PWIA are not eigenstates of the $H$.  Therefore they must have a
distribution in energy, and we may express the exact response as:
\beq
R(\qmag ,\nu) =
 \int d\nu^{\prime}F(\qmag,\nu-\nu')R_{PWIA}(\qmag,\nu^{\prime}).
\eeq
The folding function $F$ defined above,
is primarily meant to describe the width of the plane wave states. 
Scaling occurs
even in presence of interactions because this folding function
becomes independent of $\qmag$ at large $\qmag$ \cite{OB,WN82}.
Occurrence of
scaling does not imply that either the PWIA or the OSA is valid.
Fig.\ \ref{fig:ff} shows the folding function obtained from the
PWIA and exact responses for $\qmag = 10$ GeV and $\Gamma_0 = 100$
and 50 MeV.
They are calculated numerically from $R(|\bvec{q}|,\nu)$
and $R_{PWIA}(|\bvec{q}|,\nu)$ by deconvolution. 

\begin{figure}
\includegraphics[ width=275pt, keepaspectratio, angle=0, clip ]{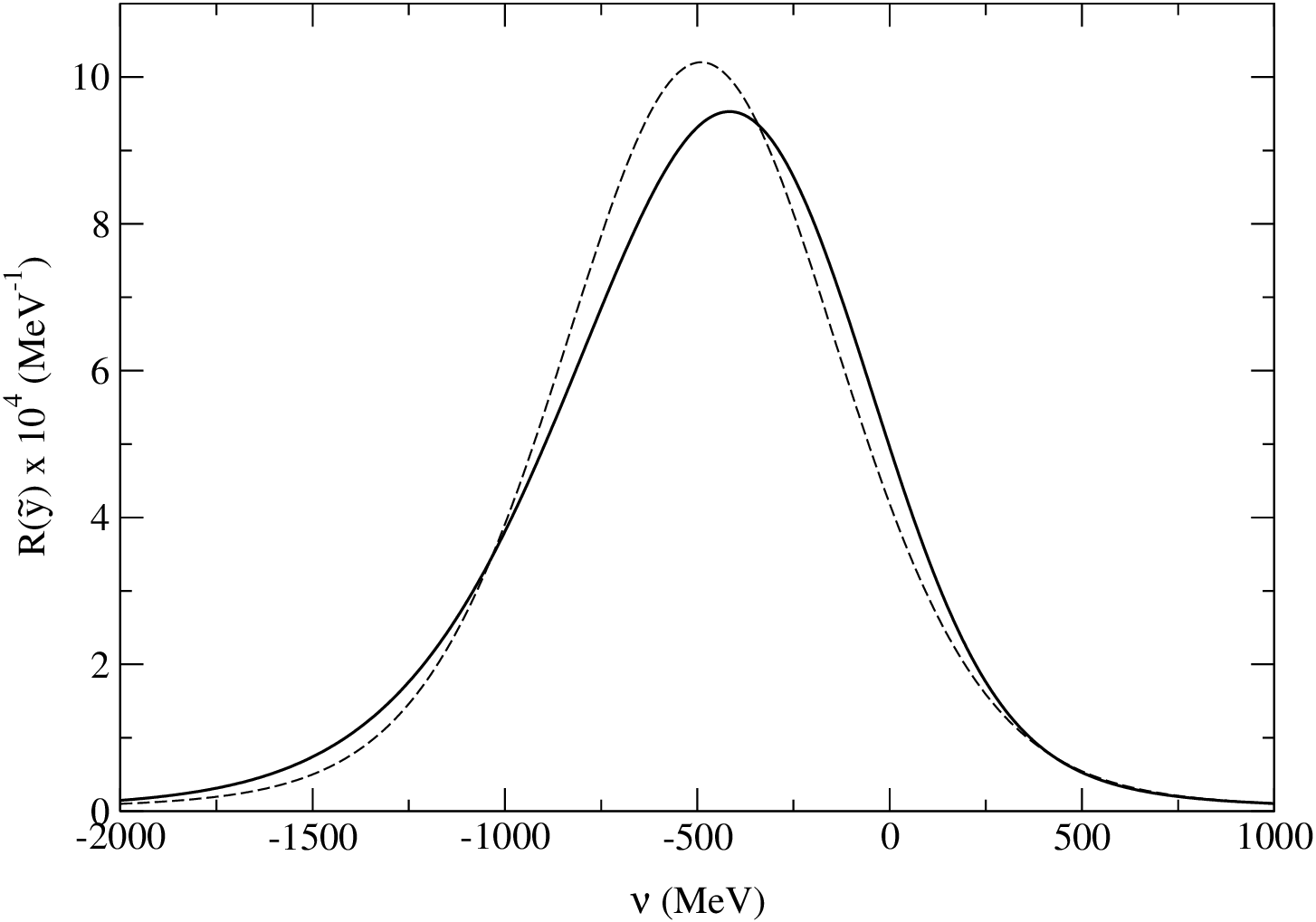}
\caption{\label{fig:gvx} The exact response (solid curve) compared
with the response obtained by folding the PWIA response with the
approximate folding function $F_A$ of Eq.(\ref{eqn:aff}) for
$\Gamma_0=100$ MeV.}
\end{figure}

In order to extract information on
the ground state wave function from the measured response,
we must first remove the FSI effects
represented by the folding function $F(\qmag,\nu)$. 
However, in most cases the folding function is not known.  In
some nonrelativistic
cases it can be calculated with quantum Monte Carlo methods
\cite{CK91}. It can also be estimated
using the Glauber approximation \cite{SS89,OB}.
Here we consider the possibility of approximating the $F$ by
a Gaussian folded with the Lorentzian distribution used in
Eq.(\ref{eqn:smear}). In this case the width of the Gaussian
is given by the interaction contribution to the exact $\Delta$
Eq.(\ref{eqn:delgs}):
\beqa
\label{eqn:deli}
\Delta^2_I &=& \gsa{V^2} - \gsa{V}^2\, \\
\label{eqn:gff}
F_G(\nu-\nu') &=& \frac{1}{\sqrt{\pi \Delta^2_I}}
e^{-(\nu-\nu')^2/\Delta^2_I},
\eeqa
and the convolution of $F_G$ with the Lorentzian is given by:
\beqa
\label{eqn:aff}
F_A(\nu-\nu')&=&
\int_{-\infty}^\infty d\nu^{\prime\prime}
F_G(\nu-\nu^{\prime\prime}) F_L(\nu^{\prime\prime}-\nu'), \\
\label{eqn:lff}
F_L(\nu^{\prime\prime}-\nu')&=&
\left(\frac{\Gamma_0}{2\pi}\right)\frac{1}{(\nu^{\prime\prime}-\nu')^2+\Gamma_0^2/4}.
\eeqa
The $F_A$ is shown in Fig. \ref{fig:ff}.  It does not have the
finer structures in the exact $F$, and it peaks at $\nu-\nu'=0$,
while the $F$ peaks at $\sim 0.2$ GeV.  Nevertheless,
a much better approximation to the response is obtained by
folding the PWIA response with the $F_A$ as shown in Fig. \ref{fig:gvx}.

In conclusion, we find that the effects of FSI, neglected in the PWIA,
become independent of $\qmag$ at large $\qmag$ for a particle
confined in a linear potential.
They increase the width of the response, and must be removed before
extracting momentum distributions or other target structure information
from the observed response.  In general one can always represent the
FSI effects by a folding function; in this simple case a Gaussian
folding function provides a reasonable approximation. Finally, the
OSA predicts a response of the wrong shape but with correct values
for the first three moments in the case of a linear potential.

\acknowledgments

The authors thank Omar Benhar and Ingo Sick for many discussions.
This work has been partly supported by the US National Science
Foundation via grant PHY 00-98353.

\bibliography{fsi}

\end{document}